\documentclass{iaus}
\usepackage{graphicx}

\title[Real Time Transient Classification] 
{Real Time Classification of Transient Events \\in Synoptic Sky Surveys}

\author[A. Mahabal et al.]   
{Ashish A. Mahabal$^1$,
C. Donalek$^1$, S.G.~Djorgovski$^{1,2}$, A.J. Drake$^1$, M.J.~Graham$^1$,
R.~Williams$^1$, Y.~Chen$^1$, B.~Moghaddam$^3$, M.Turmon$^3$}

\affiliation{$^1$California Institute of Technology, Pasadena, CA, USA\\
email: {\tt aam, donalek, george@astro.caltech.edu, ajd, mjg, roy, cheny@caltech.edu} \\[\affilskip]
$^2$Distinguished visiting professor, King Abdulaziz Univ., Jeddah, Saudi Arabia.\\
$^3$Jet Propulsion Laboratory, Pasadena, CA, USA\\
email: {\tt baback, turmon@jpl.nasa.gov}
}

\pubyear{2012}
\volume{285}  
\pagerange{1-3}
\jname{New Horizons in Time Domain Astronomy}
\editors{R.E.M. Griffin, R.J. Hanisch \& R. Seaman, eds.}
\begin{document}

\maketitle

\begin{abstract}
An automated, rapid classification of transient events detected in the modern 
synoptic sky surveys is essential for their scientific utility and effective 
follow-up using scarce resources.  This problem will grow by orders of magnitude
with the next generation of surveys.  We are exploring a variety of novel 
automated classification techniques, mostly Bayesian, to respond to these challenges,
using the ongoing CRTS sky survey as a testbed.  We describe briefly some of the methods used.
\end{abstract}


The increasing number of synoptic surveys are now generating tens to hundreds of
transient events per night, and the rates will keep growing, possibly reaching
millions of transients per night within a decade or so.
Generally, follow-up observations are needed in order to fully exploit
scientifically these data streams. In optical surveys, for
instance, all transients look the same when discovered -- a starlike object
that has changed its brightness significantly -- and yet, they could represent
vastly different physical phenomena.  Which ones are worthy of a follow-up? 
This is a critical issue for the massive event streams
(e.g., LSST, SKA, etc.), and the sheer volume requires an automated approach
(\cite{Don08}, \cite{Mah10}, \cite{Djo11a}).  

The process of scientific measurement and discovery typically operates on time scales
from days to decades after the original measurements, feeding back to a new
theoretical understanding. However, that clearly would not work in the case of
phenomena where a rapid change occurs on time scales shorter than what it takes
to set up the new round of measurements.  This results in the need for
real-time systems, consisting of computational analysis and decision engine,
and optimized follow-up instruments that can be rapidly deployed 
with immediate analysis and feedback.
These requirements imply a need for an automated classification and decision making.

The classification process for a given transient involves:
(1) obtain available
contextual archival information, and combine it with the measured parameters
from the discovery pipeline,
(2) determine (relative?)
probabilities or likelihoods of it belonging to various classes of transients,
(3) obtain
follow-up observations to best disambiguate competing classes, 
(4) use them as a
feedback and repeat for an improved classification. We describe below a few 
techniques that help in this process.
Our principal data set is the transient event stream from the
Catalina Real-time Transient Survey (CRTS; http://crts.caltech.edu; 
\cite{Dra99}, \cite{Djo11b}, \cite{Mah11}),
but the methodology we are developing is more universally applicable.

\noindent {\bf Bayesian Networks:}
Generally, the available data for any given event would be heterogeneous and
incomplete.  That is difficult to accommodate in the standard machine-learning
feature vector approach, but it can be naturally accommodated in a Bayesian
approach, such as Bayesian Networks (BN) (\cite{Mah08}).

We have used three colors obtained from the Palomar 60-inch telescope from 
follow-up observations of CRTS transients, and
two contextual parameters: Galactic latitude and proximity to a galaxy.   
Priors for six
classes have been used: CVs, SNe, Blazars, other AGN, UV Ceti, and the ``Rest''
(everything else).  We are currently adding more parameters and classes.
About 300 objects each have been used for SN and CV, and $\sim$100 for
blazars. The number statistics for other AGN and UV Ceti are still too small.
82\% of the objects classified as SN are indeed SN (79\% for CVs,
69\% for Blazars). The contamination is $\sim10 - 20\%$. Given the fact that a single
set of observations accomplished this, the potential for extending the BN, 
and combining its output with other techniques is very promising.

\noindent {\bf Lightcurve Based Classification:}
Structure in a sparse and/or irregular light curves (LC) 
can be exploited by automated classification algorithms. This can be done by
collecting LCs for different objects belonging to a class and
representing and encoding the characteristic structure probabilistically in the
form of an empirical probability distribution function (PDF). This can then be used
for subsequent classification of a LC with even a few epochs. Moreover, this
comparison can be made incrementally over time as new observations become
available, with the final classification scores improving with
each additional set of observations. This forms the basis for a real time
classification methodology.  Since the observations come in the form of flux at
a given epoch, for each point after the very first one we can form a ($\delta
m, \delta t$) pair. We focus on modeling the joint distribution of all such
pairs of data points for a given LC.  By virtue of being increments, the
empirical probability density functions of these pairs are invariant to
absolute magnitude and time shifts, a desirable feature.
Upper limits can also be encoded in this methodology,
e.g., forced photometry magnitudes at a SN location in images
taken before the star exploded.  We currently use smoothed 2D histograms to
model the distribution of elementary ($dm,dt$) sets.
In our preliminary experimental
evaluations with a small number of object classes (single outburst like SNe,
periodic variable stars like RR Lyrae and Miras, as well as stochastic
variables like blazars and CVs) we have been able to show that the density
models for these classes are potentially a powerful method for object
classification from sparse/irregular LC data.

Currently we are using the ($dm,dt$) distributions for classification in a binary
mode i.e.  successive two-class classifiers in a tree structure 
SNe are first
separated from non-SNe (the easiest bit, currently performing at a $\sim$99\%
completness), then
non-SNe are separated into stochastic versus non-stochastic variables, and then each
group further separated into more branches. The most difficult so far has been
the CV-blazar node (based on just the ($dm,dt$) density i.e., without bringing in
the proximity to a radio source since we are also interested in discovering
blazars that were not active when the archival radio surveys were done).
Currently this classifier is performing at a $\sim$71\% completness.  
We are also exploring Genetic Algorithms to
determine the optimal ($dm,dt$) bins for different classes. This will in turn
help optimise follow-up observing intervals for specific classes; see, e.g., 
\cite{Mah11} or \cite{Djo11a}.

\noindent {\bf Incorporating the Contextual Information:}
Contextual information can be highly relevant to resolving competing
interpretations: for example, the light curve and observed properties of a
transient might be consistent with it being a cataclysmic variable star, a
blazar, or a SN.  If it is subsequently known that there is a galaxy in
close proximity, the SN interpretation becomes much more plausible.
Such information, however, can be characterized by high uncertainty and
absence, and by a rich structure: if there were two galaxies nearby instead of
one then details of galaxy type and structure and native stellar populations
become important, e.g., is this type of SN more consistent with being in
the extended halo of a large spiral galaxy or in close proximity to a faint
dwarf galaxy?  The ability to incorporate such contextual information in a
quantifiable fashion is highly desirable.  We have been compiling priors for
such information as well. These then get incorporated into the Bayesian network
mentioned earlier.

We are also investigating the use of crowdsourcing (citizen science) as a
means of harvesting the human pattern recognition skills, especially in the
context of capturing the relevant contextual information, and turning them into
machine-processable algorithms.  A methodology employing contextual knowledge
forms a natural extension to the logistic regression and classification methods
mentioned above. This will be necessary for larger future surveys where the
data flow will exceed the available human resources, and moreover, it would
make such classification objective and repeatable.  It also represents an
example of a human-machine collaborative discovery process.

Transients can also be
found using the technique of image subtraction using a matched 
older observation, or a deeper co-added image
(\cite{Dra99}). If the images are properly
matched, transients stand out as a positive residual. 
When used with white light as is the case with CRTS, the difference
images tend to have bipolar residuals thus leading to false detections.
We have been experimenting with these to look for
supernovae in galaxies using citizen science where a few amateur astronomers
regularly look at the galaxy images along with the residuals presented to them
A large number of SNe have been found in
this fashion (see \cite{Pri11} for an example, and http://nesssi.cacr.caltech.edu/catalina/current.html
for a list).  


A given classifier may not be optimal for all classes, nor to all types of inputs.
That is the primary reason why multiple types of classifiers have to be employed
in the complex task of classifying transients in real time.
Presence of
different bits of information can trigger different classifiers. In some cases more
than one classifier can be used for the same kinds of inputs.
An essential task, then, is to derive an optimal event classification, given inputs
from a diverse set of classifiers such as those described above.
Combining different classifiers with different
number of output classes and in presence of error-bars is a non-trivial task
and is still under development.

{\it Acknowledgments:}  This work was supported in part by the NASA grant
08-AISR08-0085, and the NSF grants AST-0909182 and IIS-1118041.


\begin{thebibliography}{}

\bibitem[Djorgovski et~al. 2011a]{Djo11a}
{Djorgovski}, S.G., et al. 2011a, {\it Stati. Anal. Data Mining}, ref. proc. CIDU
2011 conf., eds. A. Srivasatva \& N. Chawla, in press.

\bibitem[Djorgovski et~al. 2011b]{Djo11b}
{Djorgovski}, S.G. et~al., 2011b, {\it The First Year of MAXI: Monitoring
Variable X-ray Sources},
eds. T. Mihara  \& N. Kawai, Tokyo: JAXA Special Publ., in press

\bibitem[Donalek et~al. 2008]{Don08}
{Donalek}, C. et~al., 2008, AIPC, 1082, 252, {\it Classification and Discovery in
Large Astronomical Surveys} ed. Bailer-Jones

\bibitem[Drake et~al. 1999]{Dra99}
{Drake}, A.J. et~al. 1999, ApJ, 521, 602

\bibitem[Drake et~al. 2009]{Dra09}
{Drake}, A.J. et~al. 2009, ApJ, 696, 870

\bibitem[Mahabal et~al. 2008]{Mah08}
{Mahabal}, A.A. et~al. 2008, AN, 329, 3, 288

\bibitem[Mahabal et~al. 2010]{Mah10}
{Mahabal}, A.A. et~al. 2010, ASPCS, 434, 115, {\it ADASS XIX} eds. 
Y. Mizumoto, K.-I. Morita \& M. Ohishi
 
\bibitem[Mahabal et~al. 2011]{Mah11}
{Mahabal}, A.A. et~al. 2011, BASI, 39,387

\bibitem[Prieto et~al. 2011]{Pri11}
{Prieto}, J. et~al. 2011, ApJ, Submitted (arXiv:1107.5043)


\end{thebibliography}
\end{document}